\documentclass[structabstract]{aa}  
\usepackage[varg]{txfonts}
\usepackage{longtable}
\usepackage{graphicx}
\usepackage{epsfig}
\usepackage{lscape}
\usepackage{txfonts}
\begin{document}
   \title{A new technique to efficiently select Compton--thick AGN}


   \author{P. Severgnini,
          \inst{1}
	  A. Caccianiga,
          \inst{1}
	  \and
	  R. Della Ceca
          \inst{1}
          }

   \institute{INAF-- Osservatorio Astronomico di Brera, Milano\\
              \email{paola.severgnini@brera.inaf.it}\\
             }

   \date{Received November 2011; accepted April 2012}

 
  \abstract
   {}
  {We present a new efficient diagnostic method, based on mid--infrared and
X--ray data, to select local (z$<$0.1) Compton--thick AGN with the aim of 
estimating their  surface and space density.}
{We define a  region in the X--ray-to-mid--IR [F(2-12
keV)/F$_{25}$$\nu$$_{25}$] vs. X--ray color (HR) plane associated to
Compton--thick AGN, i.e.  [F(2-12 keV)/F$_{25}$$\nu$$_{25}$]$<$0.02  and
HR$>$-0.2. 
On the basis of this selection method we build up a sample of 43 Compton--thick
AGN candidates using data from IRAS Point Source Catalog and 2XMM-Newton catalogues.
In order to test the efficiency of the proposed method in selecting
Compton--thick AGN we use the results of the X--ray spectral analysis performed
on all the sources of our sample (presented in a parallel work). After
taking into account the different selection effects, we have estimated the number of
Compton--thick in the local Universe and their density down to the IRAS flux
limit of F$_{25}$=0.5 Jy.}
   { We find that the diagnostic plot proposed here is an  efficient method to
select Compton-thick AGN in the nearby Universe since $\sim$84\%  of the sources populating the proposed Compton--thick
region are actually  Compton--thick AGN. Twenty percent are newly-discovered
Compton--thick AGN.
We then estimate  the surface density  of Compton--thick AGN down to the IRAS
PSC catalogue  flux limit (F$_{25}$= 0.5 Jy) that turns out to be $\rho^{CT}
\sim 3*10^{-3}$ src~deg$^{-2}$. 
After estimating an equivalent IR--to--hard--X--ray limiting flux, we
compare our result with those found with
SWIFT--BAT. We find that the surface density derived here is a factor 4 above the  density
computed in the hard X--ray surveys. This
difference is ascribed, at least in part, to a significant contribution ($\sim$60--90\%) of the
star--forming activity to the total 25$\mu$m emission  for the sources in our
sample.  By  considering only the 25$\mu$m AGN
emission, we estimate a surface density  of  Compton--thick AGN
which is consistent with the results found  by hard X--ray surveys.
 Finally, we estimated the co-moving space density of Compton--thick AGN
with intrinsic L$_X$$>$10$^{43}$ erg s$^{-1}$ (0.004$<$z$<$0.06):
$\Phi$$_{C-thick}$$\sim$(3.5$^{+4.5}_{-0.5}$)$\times$10$^{-6}$ Mpc$^{-3}$.  
The prediction  for Compton--thick AGN  based on the
synthesis model of X--ray background in Gilli et al. (2007) is consistent 
with this value.
}
   {}


   \maketitle %

\section{Introduction}

A complete knowledge of the local Active Galactic Nuclei (AGN) demography (i.e.
their census and physical properties) is the essential starting point to be able
to study the AGN evolution at cosmological distances. Indeed  all models developed so far to address the problem of birth and
growth of Super Massive Black Holes (SMBHs) in galaxies  are forced to
reproduce  many observational constraints among which the correct mass and
number of AGN observed locally (Marconi et al. \cite{Marconi}). While unobscured AGN can be easily detected and studied both
in the optical band and in X--rays, the detection of absorbed AGN becomes more
and more difficult as the amount of circum-nuclear obscuring medium intercepted
along the line of sight increases. This is particularly true for heavily
obscured sources (intrinsic column density, N$_H$$>$5 $\times$10$^{23}$
cm$^{-2}$) and even more for 
Compton-thick AGN (N$_H$$>$10$^{24}$ cm$^{-2}$) that are predicted to
constitute more than half of the total number of AGN (Gilli et al. \cite{Gilli}). While for less obscured AGN the X-ray photons above few
keV can penetrate the torus making the source nucleus, at least partially,
directly visible to the observer and the column density and luminosity
measurable, for Compton-thick AGN the primary radiation is almost
completely absorbed in the X--rays. For these sources,   the spectrum below 10
keV, is dominated by the so called Compton reflection/scattering component
(e.g. continuum emission reflected by the torus) which is more than an order of
magnitude fainter with respect to the  direct component. Moreover, in spite of
the different values of intrinsic N$_H$, the shape of Compton-thin and
Compton-thick AGN spectra below 10 keV could be very similar. Indeed, if the
statistics is not really good enough, this part of the spectrum could be usually well
fitted by an absorbed  (N$_H$$\sim$5$\times$10$^{23}$  cm$^{-2}$) transmitted
component or by a Compton reflection component  (see e.g. Maiolino et al.
\cite{Maiolino98},
Braito et al., \cite{Braito04}).  Because the reflection component has  a broad Compton
reflection hump in the 15--100 keV continuum, harder data are important to
complement lower energies data and to investigate the nature of the sources
(e.g. Severgnini et al. \cite{Severgnini11}, Trippe et al. \cite{Trippe}).

Even if the absorption is less severe above 10 keV, nonetheless
even harder X--ray surveys could be strongly biased against the selection of
Compton--thick AGN due to the Compton down-scattering effect (Matt et al.
\cite{Matt}, Malizia et al. \cite{Malizia}, Burlon et al. \cite{Burlon}). In particular, by using a complete sample of AGN detected by SWIFT--BAT in the first
three years of the survey, Burlon et al. (\cite{Burlon}) have shown and quantified these
effects at energies higher than 15 keV for mildly ($N_\mathrm{H}$ of the order
of a few times 10$^{24}$ cm$^{-2}$) and heavily
($N_\mathrm{H}$$\geq$10$^{25}$ cm$^{-2}$) Compton-thick AGN. They estimate that for a mildly
Compton-thick AGN only 50\% of the nuclear trasmitted flux is visible above 15
keV and this fraction become only a few percent  for heavily  Compton-thick AGN.
Therefore, even using hard X--ray data, Compton thick
sources are very difficult to detect and the computation of their volume density, requires significant corrections.

An alternative wavelength  for the selection of heavily obscured AGN is the
mid--InfraRed (mid--IR) band (see e.g. Georgantopoulos et al. 
\cite{Georgantopoulos} and references therein), where the optical and UV photons of the primary source is
re-emitted after having been reprocessed by hot dust. Since this band is less 
affected by obscuration than optical band and X-rays, AGN selection at these
wavelengths is less biased against obscured AGN. 
However, AGNs usually represent only a small fraction of all  
sources detected in IR surveys compared to the far
more numerous  IR emitters, such as galactic sources and normal and
starburst galaxies. For this reason, to efficiently select AGN, it is convenient to
complement the IR band with X--ray data, where the galaxy and star contribution is
minimal. By comparing 2-10 keV and IR fluxes it is possible to distinguish
unobscured from obscured sources being the first one relatively unbiased with respect
the extinction, while the second one strongly depressed as the 
N$_H$ value increases.

In this paper we present a well defined sample of Compton--thick AGN  selected
in the local Universe by combining mid--IR (IRAS) and X--ray (XMM--Newton) data.
The method/diagram used to select the sample is discussed in Sect.~2 and the
sample is presented in  Sect. 3. We discuss the efficiency and  the completeness
of the method. We derive the Compton-thick AGN surface density in Sect.~4 and their
space density in Sect.~5 where we also compare our results  with those found in the
literature.  Summary and conclusions are presented in Sect.~6.

\section{Diagnostic plot}

As already mentioned, one way to select heavily obscured AGN  candidates and to
distinguish them from less obscured sources is to compare the X--ray emission
below 10 keV (strongly depressed by the absorption in Compton-thick AGN) with
the emission from other bands less affected by the absorption, like harder
X--rays or mid--IR  (12--25 $\mu$m) band  (produced by the presence of
large amounts of dust absorbing, thermalizing and re-emitting the optical and UV
photons of the primary source). While hard X-rays can be strongly affected by
the Compton down scattering effect, mid--IR selection appears to be relatively
unbiased with respect to extinction even in  the case of Compton-thick sources
(Brightman \& Nandra \cite{Brightman};  Horst et al. \cite{Horst}).

   \begin{figure}[h]
   \centering
   \includegraphics[angle=0,width=9cm]{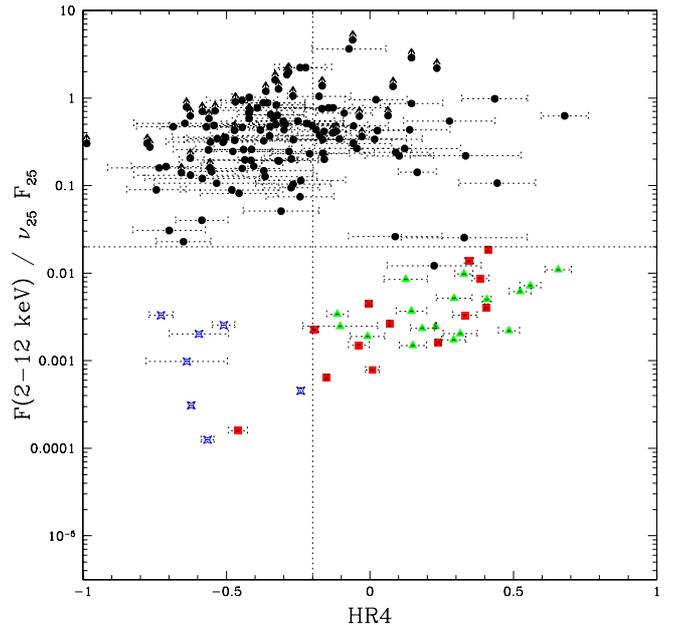}
      \caption{F(2-12 keV)/($\nu_{25}$F$_{25}$) vs. HR4 diagnostic plot. Filled circles
(black symbol in the electronic version only) are unabsorbed and absorbed
Compton-thin AGN (N$_H$$<$10$^{24}$ cm$^{-2}$) taken from two different X--ray
samples in the literature (XMM--HBS sample - Della Ceca et al. \cite{Rdcb}; XMDS survey -
Tajer et al. \cite{Tajer}; Polletta et al. \cite{Polletta07}). Stars (blue objects in the electronic version
only) are a sample of local
star-burst galaxies (Ranalli et al. \cite{Ranalli}) and squares (red objects in the electronic version only) and
triangles (green objects in the electronic version only) are local "confirmed" and "candidate" Compton-thick AGN, respectively, taken by the
compilation of Della Ceca et al. (\cite{Rdca}).}
         \label{}
   \end{figure}

Starting from this consideration, we propose here a new diagnostic plot to
select Compton--thick AGN in the local Universe. This plot is based on the
combination of the ratio between the 2-12 keV (F(2-12 keV)) and the mid--IR
(F(mid-IR)) flux  with the XMM--Newton colors (hardness ratio HR). We expect
that  Compton--thick sources are characterized by a lower F(2-12 keV)/F(mid-IR)
ratio with respect to less  obscured AGN (see e.g. Polletta et al. \cite{Polletta06},
Severgnini et al. \cite{Severgnini08}). Since starburst galaxies are characterized by similarly
low values of F(2-12 keV)/F(mid-IR) ratio, we propose here to use the X--ray 
colors to separate star-forming galaxies from Compton--thick AGN. While obscured
AGN are characterized by hard X-ray emission, the soft emission due to
star-formation activity will produce lower HR values (i.e. HR$<$-0.1,  see Della
Ceca et al. \cite{Rdc04}) with respect to that of obscured AGN.

As a first step we have plotted the X--ray and mid-IR information for different 
samples of X--ray sources for which the nature has been already studied in  the
literature (i.e. unabsorbed and absorbed Compton-thin AGN; Compton-thick AGN and
star--forming galaxies). The diagram is shown in Fig. 1 where the F(2-12
keV)/($\nu_{25}$F$_{25}$) is plotted as a function of  HR4\footnote{HR4 is
defined using the two following bands: 2-4.5 keV and 4.5--12 keV:
HR4=$\frac{CTS(4.5-12 keV) - CTS(2-4.5 keV)}{CTS(4.5-12 keV) + CTS(2-4.5 keV)}$,
where CTS are the vignetting corrected count rates in the energy ranges reported
in bracket.  See Watson et al. (2009) for details.}.  We use this figure to
define the region where looking for Compton--thick AGN: F(2-12 keV)/($\nu_{25}$
F$_{25}$)$<$0.02 and HR4$>$-0.2. Filled black circles (131 objects) are all the
sources with mid--IR information\footnote{For these sources, 24 $\mu$m
Spitzer/MIPS data have been used.  In order to adopt an uniform notation for all
the sources in the paper, we report in Fig.~1 the 25 $\mu$m fluxes, assuming a
negligible correction to go from 24 and 25 $\mu$m flux in $\nu$F$_{\nu}$.}
belonging to two X--ray  different surveys: the XMM-Hard Bright Sample
(XMM--HBS, Della Ceca et al. \cite{Rdcb}, Caccianiga et al. \cite{Caccianiga04},
Severgnini et al. \cite{Severgnini08}) and  the XMDS survey  (Tajer et al.
\cite{Tajer}, Polletta et al. \cite{Polletta07}).  X--ray information have been
taken from  the 2XMM--slim catalogue (Watson et al. \cite{Watson}). The
XMM--HBS source plotted in Fig. 1 are sources for which we obtained Spitzer
proprietary data (cycle-3, P.I. Severgnini); they have a redshift range of
0.1$<$z$<$0.7. The XMDS sources are mainly at z$<$1.5 with some
sources up to z=3.5 (see redshift distribution in Tajer et al. \cite{Tajer}). 
All but one  (the filled circle in the bottom part of the panel, F(2-12
keV)/($\nu_{25}$F$_{25}$)=0.012, HR4=0.23) have F(2-12
keV)/($\nu_{25}$F$_{25}$)$>$0.02 (see  Fig. 1) and for all of them there is no
evidence for the presence of a Compton-thick AGN  (see the  relevant papers).
The only source in which a Compton--thick AGN could be present is the filled
circle in the bottom part of the panel, see Polletta  et al.
(\cite{Polletta07}). Stars (7 sources, blue objects in the electronic version
only) are  local optical selected  star-forming galaxies taken  from the sample
of  Ranalli et al. (\cite{Ranalli}). We have considered only those sources
without evidence of a possible AGN. Finally, squares (13 sources, red objects in
the electronic version only) and triangles (17 sources, green objects in the
electronic version only) are local (z$<$0.05) ``confirmed" and ``candidate"
Compton-thick AGN, respectively, taken from the compilation of Della Ceca et al.
(\cite{Rdca}).  The so called ``confirmed" Compton--thick have been  identified
thanks to observations above 10 keV with BeppoSAX, INTEGRAL, SWIFT/BAT and
SUZAKU, while the  ``candidate" Compton-thick AGN are sources with observations
only below 10 keV.  Both for star--forming and for Compton--thick AGN we have
considered only those sources  present in the 2XMM--slim catalogue (Watson et
al. \cite{Watson}) and with an IRAS detection. All but one (NGC~3690, see
Section~4) of the local Compton--thick plotted in Fig.~1 are placed in the
lower-right part of the diagram. 

Even if the comparison of different samples, selected in different
ways and with different redshifts, is not indicative  of the real
efficiency and completeness of the proposed method, at first glance,  it suggests that this
diagram could be actually reliable in selecting local  Compton-thick AGN. 

In the next section we will test the efficiency of the proposed  method and we
will investigate if this diagram can provide a well defined and complete  sample of
local Compton-thick AGN from which it is possible to derive their surface and space
density.

\section{The sample of Compton--thick AGN candidates}

To build up a new sample of Compton--thick candidates using the diagram shown
in  Fig. 1, we have cross-correlated the IRAS Point Source Catalog 
(PSC, 245889 sources, see 
http://irsa.ipac.caltech.edu/IRASdocs/exp.sup/index.html for details) at
25$\mu$m (we exclude sources with a 25 micron 
flux density quality flag equal to 1, corresponding to upper limit, see Helou \& Walker \cite{Helou}) with
the incremental version of the v1.0  2XMM slim catalogue that contains 221012
sources.  We consider only sources
with F(4.5-12keV)$>$10$^{-13}$ erg cm$^{-2}$ s$^{-1}$ and  likelihood
parameter $>$ 12 in the 0.2--12 keV band in order to maximize the number of
counts for each source and to perform a reliable spectral analysis. To
minimize the possible contamination of Galactic sources, we select only those
sources having a high Galactic latitude ($\mid$b$^{II}$$\mid$$>$20$^\circ$).
We have used a  matching radius of 15$\arcsec$ (see
http://heasarc.gsfc.nasa.gov/W3Browse/iras/iraspsc.html) and a second step we
have excluded all the sources having an X--ray counterpart more than
10$\arcsec$ (see Watson et al. \cite{Watson}) away from the optical source
associated to the infrared emission reported in the PSC catalogue. By
repeating several times the same correlation by shifting in declination one of
the two catalogues of several degrees, we find that the number of spurious
sources is negligible ($<$1). The final list contains  145 IRAS(25$\mu$m)-2XMM
matches with  a mid--IR flux at 25 $\mu$m ranging from 0.14 to 544 Jy.

  \begin{figure}[t]
   \centering
   \includegraphics[angle=0,width=9cm]{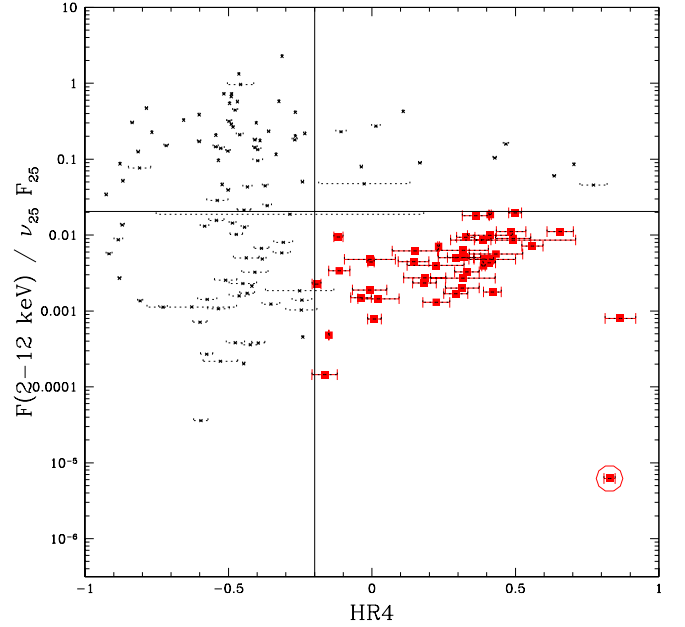}
   
      \caption{F(2-12 keV)/($\nu_{25}$F$_{25}$) vs. HR4
      diagnostic plot for the 145 source 
      found by cross-correlating the PSC 
      IRAS catalogue at 25$\mu$m and the 2XMM catalogue. Filled squares (red symbols in
      the electronic version only) are the 44
      sources that have flux ratios and X--ray colors typical of Compton--thick 
      AGN. The isolated object in the bottom--right part of the diagram
      marked with an empty circle is the only Galactic source (V* R Aqr) 
      present in the Compton--thick candidate region.}

         \label{}
   \end{figure}
   
  \begin{figure}[h!]
   \centering
   \includegraphics[angle=0,width=9cm]{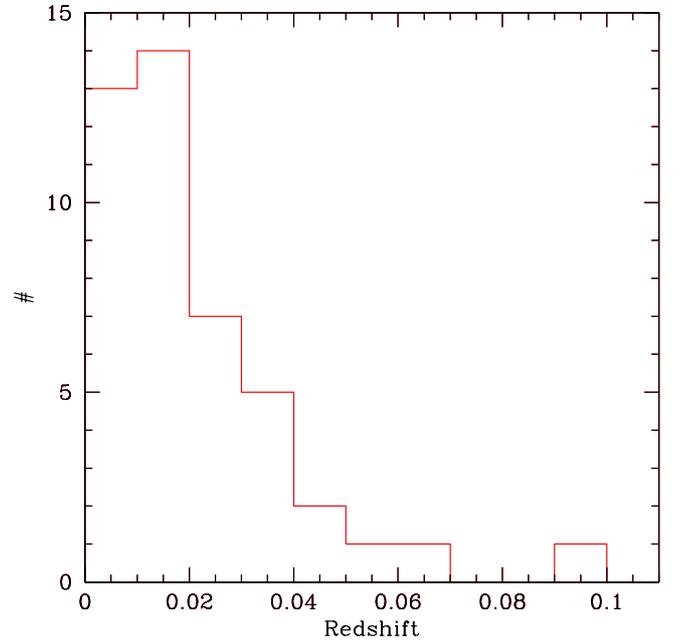}
      \caption{Redshift distribution of the 43 extragalactic sources that 
      lie in the Compton--thick region of the plot reported in Fig.~2.}
         \label{}
   \end{figure}

As discussed in the previous section, on the basis of the plot reported in Fig. 1 we define as
heavily obscured AGN region the  zone with Fx/($\nu_{IR}$ F$_{IR}$)$<$0.02 and HR4$>$-0.2
(i.e. the lower--right region). By plotting the results of the IRAS-2XMM cross-correlation on the
F(2-12keV)/($\nu_{25}$F$_{25}$)--HR4 plane we find 44 sources in the region associated to
heavily obscured AGN (see Fig. 2, filled squares, red symbols in the electronic version only), 43 of which are extragalactic sources (the
only Galactic object is the isolated encircled source  in the bottom right part of the
diagram).  For all sources, the redshift is already reported in the literature (see Table~1).
The redshift distribution is shown in  Fig. 3. The full sample is at z$<$0.1 and 98\% of the
sources have z$<$0.07.

\subsection{X--ray properties of the Compton--thick candidates}

As a first step, we checked in the literature if an X--ray classification 
exists for all the sources in Table~1.  We  find that a large fraction of them
(30/43) are already known as Compton-thick AGN on the basis of a direct measure
of the absorption cut--off or  through indirect arguments, such as the presence of a
strong Iron emission line at 6.4 keV. Twenty Compton-thick belong to the compilation of Della
Ceca et al. (\cite{Rdca}) and they are plotted also in Fig. 1.
Five are known as
Compton--thin AGN.  For 13 Compton-thick AGN, the classification has been
confirmed also thanks to observations above 10 keV  (NGC 424, NGC 1068, Mrk 3,
NGC 3079, Mrk 231, M51, NGC 6240, NGC 7674, see Della Ceca et al. \cite{Rdca} -
Mrk 273, see Teng et al. \cite{Teng} - NGC 2273, see Awaki et al. \cite{Awaki} -
NGC 7582, see Bianchi et al. \cite{Bianchi09} - NGC 1365, see Risaliti et al.
\cite{Risaliti05} - AM 1925-724, see Braito et al. \cite{Braito09}). Three of
them (NGC 1365, NGC 7674 and NGC 7582) show rapid Compton-Thick/Compton-Thin
transitions and they are known as "changing--look" AGN.  Finally, one source is
Arp~220, that has been extensively studied so far in several bands.  Many of the
features detected in the X--ray spectrum (a flat continuum - Ptak et al.
\cite{Ptak} - and a  prominent Fe K$\alpha$ emission line, EW$\sim$1.9 keV -
Iwasawa et al. \cite{Iwasawa}) suggest the  presence of a heavily obscured AGN.
This hypothesis is also the most favorite one after the analysis of the Suzaku
data by Teng et al. (\cite{Teng}).  Thus we consider this object as a possible
Compton--thick AGN.

In order to obtain a uniform analysis for all of the Compton--thick AGN
candidates, we have performed our own spectral analysis using the XMM data for
the 42 sources  in the sample with more than 100 net counts in the 0.5-12 keV
range. For the remaining  one (NGC 5879) the statistics of the XMM data is
not good enough (from 15 to 60 counts) to allow an appropriate X--ray spectral
analysis. Since  good ($>$100 net counts) {\it Chandra} data are available for
this latter object, we
use them to study its X--ray spectral properties. We have applied both disk
reflection models (i.e. {\it pexrav} model  Magdziarz \& Zdziarski
\cite{Magdziarz}) and the recent model proposed in the case of  neutral toroidal
X-ray re-processor in AGNs (Murphy \& Yaqoob \cite{Murphy}).  We inferred the
possible presence of  Compton-thick AGN mainly through the detection of the
absorption  cut--off or through indirect arguments, such as the presence of 
dominant 2-10 keV reflection/scattering emission plus a prominent (EW$>$400 eV)
Iron line. To further investigate the nature of our sources we have
obtained hard X--ray data from the catalogue obtained after 54 months of
SWIFT--BAT observations (Cusumano et al. \cite{Cusumano}) for 17 sources and we
used this X--ray hard data to better constrain the absorbing column density. We
also obtained Suzaku observations for two of them  (IRAS 04507+0358 and
MCG-03-58-007, 100 ksec each).  A detailed description of the analysis done on
the XMM data of the sources not known as Compton--thick from the literature,
combined, in some cases, with BAT and Suzaku data, will be  reported in a
companion paper (Severgnini et al. in prep.).   

In the last two columns of  Table~1 we report:  the  satellites/instruments
from which we have taken the X--ray data and the  X-ray classification. We
classify a source as "Compton--thick" AGN (32 sources) if we obtain an
indication  of a column density (N$_H$) larger than 10$^{24}$ with both the
models used (disk--reflection and toroidal models), while we adopt the
classification "Compton--thick?"  for 3 sources for which the presence of a
Compton--thick AGN is model dependent and in the case of Arp~220. Our X--ray
analysis confirms the classification as Compton--thick AGN taken from the
literature in all cases except for one source  (IC~4995, see Guainazzi et al.
\cite{Guainazzi}). In addition to these, we find 7 newly discovered
Compton--thick AGN (marked with a double asterisk in Table~1). One of these
is  IRAS 04507+0358,  that we have extensively discussed in Severgnini et al.
(\cite{Severgnini11}).

\subsection{Efficiency and completeness of the method}

{\bf Efficiency -} The diagnostic plot proposed here could be considered as an
efficient way to select local Compton-thick sources in the nearby Universe. As
reported in the previous section, for $\sim$84\% of the sources populating
the  Compton--thick region  the presence of a Compton--thick AGN is suggested or
confirmed by the X--ray spectral analysis. For comparison, the efficiency in finding Compton--thick AGN
using other X--ray--to--mid--infrared diagnostic ratio (e.g. L$_X$/L$_{6
micron}$, as recently re-proposed by Georgantopoulos et al.
\cite{Georgantopoulos}) is 50\% and in a hard X--ray survey, like that  presented
in Burlon et al. (\cite{Burlon}) or in the recently published  all-sky sample of AGN
detected by BAT in 60 months of exposure (Ajello et al. \cite{Ajello}),
is about 5--6\%.

~\\ 

{\bf Completeness -}   While the samples reported in Fig. 1 can
not be used to estimate the  efficiency of the proposed method, we can use them
to state, at first glance, its completeness. Indeed, even if the Compton--thick
sample doesn't include all the Compton-thick AGN known so far, the sources
plotted in Fig.1  have been not chosen on the basis of their X--ray--to--IR
ratio or on the basis of their X--ray colors. In this sense, they can be considered
representative of the AGN Compton--thick population.

As already discussed in Section 2, there is just 1 source, NGC
3690, in the Compton--thick  compilation reported by Della Ceca et al.
(\cite{Rdca}) that fall outside the Compton--thick region considered here. 
We discuss it in more details in the following.

NGC~3690, falls in the lower-left part of the plot (i.e. the star--forming
region). It is one of the two merging galaxies of the LIRG Arp~299 (Sanders et
al. \cite{Sanders}; Heckman et al. \cite{Heckman};  Della Ceca et al.
\cite{Rdc02}; Ballo et al. \cite{Ballo}). The optical spectroscopic
classification puts this source at borderline between starburst and LINER
(Coziol et al. \cite{Coziol}), while the X--ray analysis clearly reveals the
presence of a strongly absorbed AGN in the system (Della Ceca et al.
\cite{Rdc02}; Ballo et al. \cite{Ballo}).  The 2--10 keV continuum is  due to a
combination of  reprocessed AGN emission (reflection and/or scattering) and
starburst activity which, most probably dominates and produces the soft HR4
(=-0.396) observed. This is the only source already known as
Compton--thick AGN which lies in the star--forming region of both Fig. 1 and
Fig. 2.  As a further check of the possible presence of Compton--thick AGN in
this part of the plot, we have verified how many sources of Fig. 2 placed in
this part have a detection in the hard X--rays. To this end we considered 
the 54--months SWIFT--BAT catalogue by Cusumano et al. (\cite{Cusumano}).  The
only source with hard X--ray detection is M82, which is considered one of the
prototype of starburst galaxies (Sakai \& Madore \cite{Sakai}). The hard
emission detected in this source is most probably due to the presence of a
Ultra--luminous compact X--ray source (X--1 , Miyawaki et al. \cite{Miyawaki})
with  a bolometric luminosity of (1.5--3)$\times$10$^{40}$  erg s$^{-1}$. No
evidence of Compton--thick AGN in this object and no evidence of Compton--thick
AGN in the other sources populating the star--forming region of the plot can be
derived by  hard X--ray observations. This part of the plot is populated
by star--forming galaxies or low--luminosity Seyfert/LINERs in which the X--ray emission is most
probably dominated by star--forming activity.

By taking into account that we are considering only those sources with  IRAS
PSC and XMM-Newton information and with F$_{25}$$>$0.5 Jy,
$\mid$b$^{II}$$\mid$$>$20$^\circ$ and  F(4.5-12 keV)$>$10$^{-13}$ erg
cm$^{-2}$ s$^{-1}$, there are 20 Compton--thick AGN in the compilation of
Della Ceca et al (\cite{Rdca}) that satisfy these criteria. Since, as quoted
above, our Compton--thick selection miss 1 of them, we state that, in a first
approximation, our method is complete at 95\% ({\it C}$\sim$0.95).

\section{Compton--thick AGN surface density}

We now want to use the selected sample to estimate the number of  Compton--thick
AGN in the local Universe (z$<$0.1). As discussed above, mid--IR band is less
affected by the absorption  and, therefore, the selection function is expected to
be relatively flat (see also Brightman \& Nandra 2011).  

Since the IRAS survey is complete down to $\sim$0.5 Jy  at 25 micron (Helou \& 
Walker \cite{Helou}), hereafter we will refer to this flux limit to derive
statistical  considerations on Compton--thick  AGN. Out of the 43 sources in the
Compton--thick box, 34 have F$_{25}$$\geq$0.5 Jy. Twenty-six are classified as
"Compton--thick" and 3 as "Compton--thick?".  Thus we observe 26-29
Compton--thick AGN down to a flux limit of F$_{25}$=0.5 Jy.

In order to derive the density of Compton--thick we have to take into account
three  problems that affect the sample discussed here.

First, we search for local Compton--thick AGN by considering only those sources
with F(2-12 keV)/($\nu_{25}$F$_{25}$)$<$0.02 and HR4$>$-0.2. 
We have already discussed the completeness {\it C} of our selection method in Sect. 3.2.

Second, the {\it effective} area of sky covered by our sample is  not known
a-priori. The problem  is connected to the  2XMM--{\it Newton} catalogue which 
includes both  sources falling serendipitously in the  field-of-view of the
telescope and the targets  of the pointings. Considering only the serendipitous
sources, the sky area  covered by the 2XMM--{\it Newton} catalogue is relatively
small ($\sim$360 deg$^2$,  Watson et al. \cite{Watson}). Based on previous
estimate of the surface density of Compton--thick AGN, the expected  number of
nearby Compton--thick AGN falling by chance  in this area is negligible  ($<$1,
see e.g. Burlon et al. \cite{Burlon}) so our sample is  made almost exclusively
by sources that  have been targeted by the XMM-{\it Newton}  telescope (all
but 2 sources are targets). 
Therefore,  the probability of  finding a source in the 2XMM catalogue, is not
anymore connected to the real area  covered by the catalogue but it  depends on
how frequently that type of astrophysical  source has been  observed.  Ideally,
if all or nearly all the sources under study  with a flux above a given flux
limit have been pointed by XMM-{\it Newton}, the covered  area can be considered
equal to the entire sky. If, on the contrary, only a  fraction of sources have
been pointed, the effective area must be scaled down   proportionally. We call
this fraction $F_{target}$.  Since the pointed sources do not  constitute a
representative sample, the value of $F_{target}$ is expected to be different 
for different classes of astrophysical sources. 

Third, our sample is flux limited in two different bands, i.e. the 25 micron  
and the X-ray bands, so it cannot be considered as a purely mid--IR selected 
sample. For a  given mid--IR flux limit, the effect of the X-ray limit is to
exclude a  number  of sources. We refer as $F_{Xl}$ the fraction of objects that
pass the X-ray  limit (i.e. $F_{Xl}$=1 if the X-ray limit is not important).

If all the three factors discussed above ({\it C}, $F_{target}$ and $F_{Xl}$) are
estimated,  we can infer the number of Compton--thick AGN at the IRAS flux limit
starting from the computed number of Compton--thick present in the sample
(N$_{CT}$) and the relevant surface density:

\begin{center}

N$_{CT} (F_{25}> F_{LIM}) = \frac{N_{observed CT}}{{\it C} \times F_{Xl} \times
F_{target}}$

$\rho^{CT} (F_{25}> F_{LIM}) = \frac{N_{CT}}{A_{20}}$ src $\deg^{-2}$

\end{center}

where $A_{20}$ is the total sky area at high Galactic latitude ($|b^{II}|>$20 
$\deg$)
and $F_{LIM}$ is the flux limit at 25 micron. 

In the following, we present different methods to estimate the two fractions, 
$F_{target}$ and $F_{Xl}$.

\subsection{Estimate of F$_{target}$}

As explained above, the sample of CT analyzed in this paper is made mainly by
sources that have been  chosen as targets of XMM-{\it Newton} telescope. In
order to quantify the value of F$_{target}$, i.e. the fraction of sources that
have  been pointed by XMM-{\it Newton}, we have analyzed  two samples of sources
that are in many aspects similar to the one considered here. The first one is 
the  sample of Seyfert2/CT AGN  discovered in the Swift-BAT survey (Burlon et
al. \cite{Burlon}) while the second one is the sample of Seyfert 2 included in
the {\it extended}  12 micron sample (Rush, Malkan \&  Spinoglio \cite{Rush}). 
The first one is a complete, flux--limited sample of local AGN at
$|b^{II}|>$15 $\deg$ collected by the Swift--BAT instrument in the first three
years of the survey, while the second one is a 12 $\mu$m flux--limited sample of
893 galaxies at $|b^{II}|>$25 $\deg$ from the IRAS Faint Source Catalogue
(Moshir \cite{Moshir}). Both samples are purely flux limited samples and, in
both cases, the selection is not related to the (soft) X-ray properties of
sources. Since the properties (IR fluxes, redshift)  of these sources are very
similar to the Compton--thick AGNs  present in our  sample  (indeed, the overlap
between these samples is large) it is reasonable to  assume  that the fraction
of Seyfert 2 in Swift-BAT sample or in the ``extended'' 12 micron sample that
have been observed by XMM-{\it Newton} gives a rough approximation of the value
of F$_{target}$. We have thus positionally cross-correlated these two catalogues
with the 2XMM catalogue.  We find that  50\% of the 12 micron classified as
Seyfert~2 have been pointed with XMM-{\it Newton}. Since  12 micron sample is
not spectroscopically complete (Hunt \& Malkan \cite{Hunt}, Brightman et al.
\cite{Brightman}) and since the optical elusiveness of X--ray selected AGN is a
well known critical problem (see e.g. Caccianiga et al. \cite{Caccianiga}, 
Severgnini et al. \cite{Severgnini03}, Maiolino et al. \cite{Maiolino03} and
references therein) we have estimated the fraction of XMM-{\it Newton} targets
including also  the sources classified as LINERS or  ``high far infrared'' 
sources (that potentially may contain an hidden Compton--thick  AGN). We find
that the fraction decreases to 40\%.

Finally, if we consider only the  AGN in the Swift-BAT sample with a measured
N$_H$ larger than 10$^{24}$ cm$^{-2}$ we find  a somewhat higher fraction
(63\%), although considering the small numbers (7 out of 11 sources observed
with XMM-{\it Newton}), this fraction is fully  consistent with the one found
considering all the Sy2s. We conclude that a reliable estimate of F$_{target}$
is 0.5$\pm$0.1.

\subsection{Estimate of F$_{Xl}$}

The value of F$_{target}$ computed above does not take into account the fact
that  we are considering only the sources with an X-ray flux above 10$^{-13}$
erg cm$^{-2}$ s$^{-1}$. The  presence of this limit, which has been set in 
order to make the X-ray  spectral analysis more reliable, exclude from our
sample a number of Compton--thick AGNs.  We evaluate the fraction of missing
sources by re-running the positional  cross-correlation between the 2XMM and
the IRAS catalogues without imposing  any limit on the X-ray flux. After the
exclusion  of Galactic sources and of ULX in nearby galaxies we find  53
sources in the ``Compton--thick box''  down to F$_{25}$=0.5 Jy. At fainter
X-ray fluxes the number of expected  spurious matches (negligible in  the
original sample) probably could be important. Therefore,  we have repeated several
times the same correlation by shifting in declination one of the two
catalogues of several degrees in order to get an estimate of the fraction of
spurious sources. 

We estimate a fraction of spurious matches of the order of 10\% so the actual
number of matches is about 48, i.e. 14 more sources  with respect to the
original sample (34 sources, including non CT sources) down to the same 25
$\mu$m  flux  in the Compton--thick box. We therefore estimate a value of
F$_{Xl}$ of  34/48 $\sim$0.7. We note that, following this method to
compute the F$_{Xl}$, we only consider the  X-ray sources that are present in the
2XMM catalogue. Therefore, sources  fainter than the 2XMM flux limit are not
included in this computation. It  could be argued that in this way the
fraction of sources missed because of  the X-ray limit is underestimated. This
would be true is we were considering  only serendipitous source. We recall,
however, that we are dealing with  sources that are targets of the XMM-Newton
observation. If a source has been  pointed, then it is usually
detected\footnote{We have verified that the Sy2 pointed by XMM have been actually 
detected. To do this, we have considered the 44 AGNs classified as 
"Seyfert type 2" in the XMM-Newton Master Log \& Public Archive and we 
have checked whether they appear also in the 2XMM catalogue of sources. 
We have found that 40 out of 44 objects are indeed present in the 2XMM 
catalogue and, therefore, they are detected. In the remaining 4 cases 
the source is not present in the 2XMM catalogue simply because the image 
has not been used to produce the 2XMM catalogue.} and,
therefore, present in the 2XMM  catalogue. On the contrary, if a source is
not  in the 2XMM catalogue, this  means that it has not been chosen as a
target. Therefore, the problem of the  sources that are not included in the
2XMM catalogue is already accounted for  during the F$_{target}$ step and it does
not require any further correction.

\subsection{The density of Compton--thick AGN}

Using the values of {\it C}, F$_{target}$ and F$_{Xl}$ derived above and the
number of  Compton--thick AGN  found in our  sample (26--29) down to a flux
limit of 0.5 Jy  at 25 micron, we can compute the number of Compton--thick AGN
and their surface density. We find:

\begin{center}
 
 $N_{CT} (F_{25}> 0.5 Jy) \sim$ 83$\pm$5

\end{center}

\begin{center}

$\rho^{CT} (F_{25}> 0.5 Jy)\sim 3 *10^{-3} src~deg^{-2}$
\end{center}

\section{Comparison with other samples}

As discussed in the previous section, to detect and study Compton--thick AGN
is not easy, even in the local Universe.  Often,  the low X-ray statistics or
the very high column density (N$_H$$>$5x10$^{24}$ cm$^{-2}$) prevent us from
deriving the amount of absorption along the line of sight by using
observations below 10 keV. For these sources, even at E$>$ 10 keV there is a
strong bias against the detection of very obscured sources, as recently
demonstrated by Burlon et al. (\cite{Burlon}). In particular, these authors
analyzed a complete sample of AGN detected by SWIFT--BAT in the first three
years of the survey. They estimate the bias of the BAT instrument against the
detection of Compton--thick AGN and they found that the real fraction of AGN
with N$_H$ ranging from 10$^{24}$  to 10$^{25}$ cm$^{-2}$ should be a factor
of 3--4 greater than the observed one, for a total of $\sim$40 expected
Compton--thick AGN down to a flux limit of  $\sim$10$^{-11}$ erg cm$^{-2}$
s$^{-1}$ and $\mid$b$^{II}$$\mid$$>$15$^\circ$.

It is now interesting to compare the results obtained here with those
reported  in Burlon et al. (\cite{Burlon}) or in the recent updated BAT
all--sky catalogue published by Ajello et al. (\cite{Ajello}). Given the
different selection band (IR and hard  X-rays respectively) we can compare
the two surveys only by assuming an  average hard X-ray-to-IR flux ratio
typical for AGN. This ratio must be  intrinsic, i.e. it should not include
the effect of Compton--down scattering  that reduces the hard X-ray flux. On
the basis of the Unified model of AGN (Antonucci \cite{Antonucci}), the
average intrinsic X-ray-to-IR  flux ratio can be simply computed using the
type~1 AGNs present in the BAT  survey of Burlon et al. (\cite{Burlon}),
since the Compton--down scattering is completely negligible at the  column
densities observed in this type of sources.  We measure an average
F$_{(15-55 keV)}/F_{25}$ ratio of  $\sim$5$\times$10$^{-11}$ erg s$^{-1}$
cm$^{-2}$ Jy$^{-1}$ which implies that the F$_{25}$(AGN)$\geq$0.5  Jy of our
surveys corresponds to an hard X-ray limit of F$_{(15-55 
keV)}\sim$2.5$\sim$10$^{-11}$ erg  cm$^{-2}$ s$^{-1}$.  Using the BAT
survey, we estimate that at this flux limit  the density  of Compton--thick
sources, corrected for the Compton-down scattering\footnote{We applied the
same  correction as estimated by Burlon et al. (\cite{Burlon}) to remove the
effect of the Compton-down scattering on the total number of Compton--thick
AGN observed in the BAT survey.}, is  7$\times$10$^{-4}$ src~deg$^{-2}$ and
8$\times$10$^{-4}$ src~deg$^{-2}$ from Burlon et al. (\cite{Burlon}) and
Ajello et al.  (\cite{Ajello}),  respectively.  These values are a factor
$\sim$4 below the density computed in  our survey (see Sect.~4.4). 

The origin of this large discrepancy could be related to the
contamination of  the observed IR flux from non-AGN activity, like the one, for
instance, due  to an intense star--formation. Indeed, a characteristic feature
of the X-ray  spectra of the CT sources in our sample is  the almost ubiquitous
presence of  a thermal component that suggests the presence of  star formation
in the host  galaxy. It is therefore possible that the observed 25$\mu$m flux
is, at least in part, due to this extra ``non-AGN'' component.
If this is the case, our sample include AGN intrinsically fainter with
respect to the Compton--thick AGN in the sample of Burlon et al.
(\cite{Burlon}).

As suggested by Fig. 1, the lower--left region in Fig. 2 should be
dominated by star--forming activity.  To evaluate the contribution  of
star--forming activity to the 25 $\mu$m emission (F$_{25}$(SF)) in addition to
the AGN (F$_{25}$(AGN)), we  have considered all the sources populating this
part of the diagram after excluding the sources classified as "Seyfert", "LINERs"
or "Star" by NED\footnote{NED (NASA/IPAC Extragalactic Database) is operated by the
Jet Propulsion Laboratory, California Institute of Technology, under
contract with the National Aeronautics and Space Administration.}. Using these  sources we can thus estimate the mean
F$_{25}(SF)$/F$_{(0.5-2 keV)}$ ratio of  the star-forming galaxies (see Fig.~4) and use it
to  estimate the F$_{25}(SF)$ in the CT AGN.  In particular, we use the
F$_{(0.5-2 keV)}$ derived from our X--ray spectral  analysis and by considering
only the 0.5--2 keV thermal component. We find that the host galaxies of our
Compton--thick AGN have 25 micron luminosities associated with the
star--formation activity that ranges from about 6$\times$10$^{8}$ L$\odot$ to
6$\times$10$^{11}$ L$\odot$ (75\% of them have L$_{25}$$<$5$\times$10$^{10}$
L$_{\odot}$), that  are in good agreement with the typical IR luminosity range
measured in local IRAS galaxies (see e.g. Rush et al. \cite{Rush}). From the 
observed  F$_{25}$ and the F$_{25}(SF)$ estimated from the soft X-ray flux we 
then obtain, by difference, the AGN contribution in all the CT sources. We find that, at the zeroth order, the mean  AGN contribution to the total 25
$\mu$m flux range from 40\% to 10\%, i.e. in our sample the galaxy contribution at
the IR band is not negligible.

  \begin{figure}[t]
   \centering
   \includegraphics[angle=0,width=9cm]{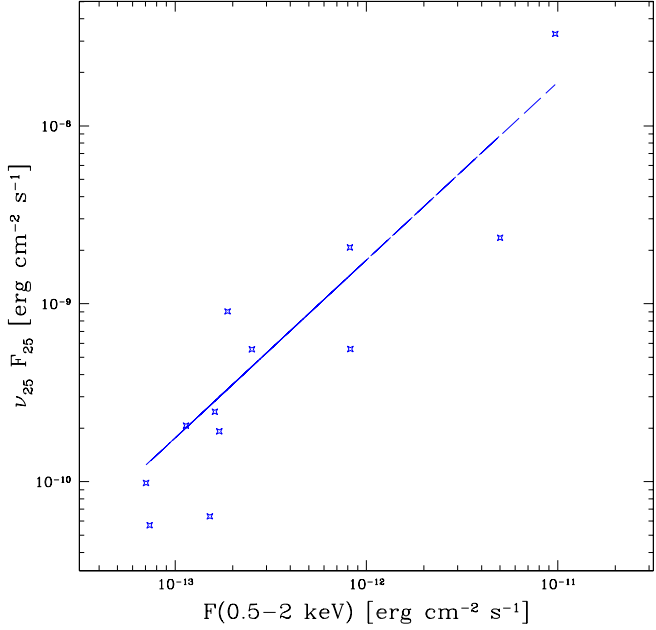}
      \caption{Mid--IR ($\nu_{25}$ F$_{25}$) vs. X--ray (F(0.5--2 keV)) fluxes of the sources populating 
      the lower--left region in Fig. 2 after excluding the sources 
      classified as "Seyfert", "LINERs" or "Star" by NED. The straight line
      (blue
      line in the electronic version only) indicate the mean value of 
      the $\nu_{25}$ F$_{25}$/F(0.5--2 keV) of these sources.}
         \label{}
   \end{figure}

We now consider only those objects in the original sample of 43 sources that  have
F$_{25}$(AGN)$\geq$0.5 Jy. We find 20 sources, 15--16 of which are  Compton--thick, i.e. the
number of Compton--thick AGN is decreased of a factor $\sim$1.8 (from 26-29 to 15--16). This
means that the Compton--thick AGN density at F$_{25}$=0.5 Jy, if only the AGN  emission is
considered, is $\sim$ (1.7$\pm$1)$\times$10$^{-3}$ src~deg$^{-2}$ a value  that, considering the
uncertainties on all the estimates, is compatible with the density estimated from the BAT survey 
(Burlon et al. \cite{Burlon}, Ajello et al. \cite{Ajello}). This  confirms that Compton down
scattering is important at hard X-ray energies and that the Compton--thick AGN densities
estimated from hard X-ray surveys must be significantly corrected, as done by Burlon et al.
(\cite{Burlon}).  Although, we have demonstrated  that the infrared band is contaminated by
star--forming emission, the corrections to apply in this case are lower with respect to those in
the hard X--rays. An IR-based selection allows the  discovery  of the majority of the sources
and, more importantly, is not  biased (in principle) against high column densities because it is
not  affected by the Compton down scattering.

We finally estimate the co-moving space density of locally Compton--thick
AGN.  In order to allow a direct comparison with recent results obtained for
higher redshift Compton--thick AGN, we estimated this density for 
Compton--thick AGN with L$_X$$>$10$^{43}$ erg s$^{-1}$.

Since we are dealing with a IR selected sample,  we consider the AGN spectral
energy  distribution (SED) reported by Shang et al. (\cite{Shang}). These
authors have compiled SED for 85 quasars using high--quality multi--wavelength
data from radio to X--rays and they constructed the median  SEDs for radio
loud and radio quiet quasars. We derive  the IR--to--X--ray luminosity ratio
for AGN on the basis of their composite SED for radio quiet. We consider all
the sources of our sample with L$_{25}$$>$4$\times$10$^{30}$ erg cm$^{-2}$
s$^{-1}$ hz$^{-1}$ (10 sources), that is the IR luminosity equivalent to L$_{(2-10
keV)}$$>$10$^{43}$ erg s$^{-1}$. After rescaling the original sample for the
different incompleteness discussed in Section 4, we estimated the co-moving
space density of local (0.004$<$z$<$0.06) Compton--thick AGN  (with the
1/V$_{max}$ method, Avni \& Bahcall \cite{Avni}): 
$\Phi$$_{C-thick}$$\sim$(3.5$^{+4.5}_{-0.5}$)$\times$10$^{-6}$ Mpc$^{-3}$  
(assuming 
H$_0$=71, $\Omega_{\lambda}$=0.7 and $\Omega_{M}$=0.3).

  \begin{figure}[t]
   \centering
   \includegraphics[angle=0,width=9cm]{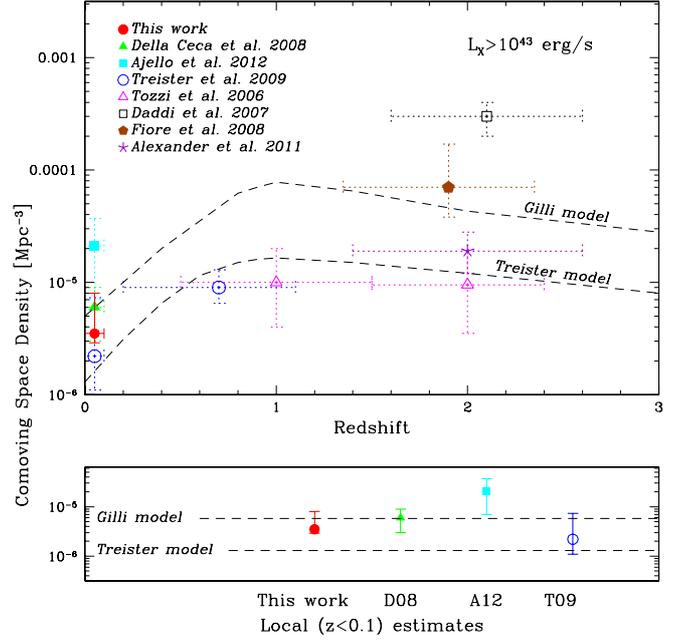}
      \caption{Co-moving space density of Compton--thick AGN. All the data and
      the model plotted in the figure  refer to L$_X$$>$10$^{43}$ erg
      s$^{-1}$. Filled circle (red symbol in the electronic version only) is the estimate obtained in
      this work, while the other local values are taken from  
      Della Ceca et al. (\cite{Rdcb}, solid triangle,
      green symbol in the electronic version),  from Ajello et al.
      (\cite{Ajello}, solid square, cyan symbol in the electronic version) and from Treister et al. 
      (\cite{Treistera}, open circle at the local redshift, blue symbol in the electronic version
      only). 
      As for higher redshift estimates: open circle (blue symbol in the electronic version only),
       filled pentagon (brown symbol in the electronic version 
      only) and open  square are the results obtained from the X--ray stacking analysis of undetected candidate
      Compton--thick AGN from Treister et al. (\cite{Treisterb}), Fiore et al. (\cite{Fiore}) 
      and Daddi et al. (\cite{Daddi}), respectively. The results obtained fron the X--ray
      spectral analysis from Tozzi et al.
      (\cite{Tozzi}) and Alexander et al. (\cite{Alexander}) are marked with  open triangles (magenta symbols in the electronic version 
      only) and star (purple symbol in the electronic version 
      only), respectively.
      The results are compared to the predictions of the models proposed by Gilli et al.
      (\cite{Gilli}) and Treister et al. (\cite{Treistera}), dashed curves.
      The local co-moving space density estimates are reported also in the lower
      panel as a function of the 
      different authors.
      }
         \label{}
   \end{figure}

In Fig. 5 we compare our estimate with the values measured by different
authors at different redshifts (open and solid symbols) and with the predictions of the  synthesis models of X--ray
background (dashed lines). In particular, our value is in good agreement with  the
co-moving space density obtained by integrating the X--ray luminosity function
of Compton--thick AGN discussed in Della Ceca et al., 2008
($\Phi$$_{C-thick}$$\sim$6$\times$10$^{-6}$ Mpc$^{-3}$ adapted to H$_0$=71) and with the estimate reported
by Treister et al. (\cite{Treistera}) for local Compton--thick AGN, while
it is lower with respect to the value reported by Ajello et al.
(\cite{Ajello}). All the data and the model reported in Fig. 5 refer to sources
with L$_{(2-10 keV)}$$>$10$^{43}$ erg s$^{-1}$. For completeness, we report in Fig. 5 also the different
estimates of the co-moving space densities for higher redshift Compton--thick
AGN  ranging from 
$\Phi$$_{C-thick}$$\sim$10$^{-5}$ Mpc$^{-3}$ to 
$\Phi$$_{C-thick}$$\sim$3$\times$10$^{-4}$ Mpc$^{-3}$ . We plot the results
obtained from the X--ray stacking analysis of undetected candidate
Compton--thick AGN from Treister et al. (\cite{Treisterb}), Fiore et al.
(\cite{Fiore}) and Daddi et al. (\cite{Daddi}) and those obtained from the 
X--ray spectral analysis from Tozzi et al. (\cite{Tozzi}) and Alexander et al.
(\cite{Alexander}).

Finally, as for the comparison with the  synthesis models of X--ray
background, the results obtained by using the model of  Gilli et al.
(\cite{Gilli}) are consistent,  within the uncertainties, with the space density
derived in this work, while the prediction obtained by the model presented in
Treister et al. (\cite{Treistera}) is lower.

\section{Summary and conclusion}

We have presented a new method to select Compton-thick AGN in the local
Universe, evaluated its efficiency and completeness.

The proposed method is based on the combination of the X--ray--to--mid--IR
flux ratio (F(2-10 keV)/($\nu_{25}$F$_{25}$)) with the X--ray colors (HR4). We
define an heavily obscured region  (F(2-10 keV)/($\nu_{25}$F$_{25})$$<$0.02
and HR4$>$-0.2) where  Compton--thick AGN are typically found. After
cross-correlating the  IRAS Point Source Catalog with the bright and hard
(F(4.5-12keV)$>$10$^{-13}$ erg cm$^{-2}$ s$^{-1}$) end of the 2XMM-Newton
catalog, we find 43 Compton--thick AGN candidates. Through a detailed X--ray
spectral analysis (presented in a companion paper) we have found that about
84\% of them are Compton--thick AGN. Twenty percent of
the selected Compton--thick are newly-discovered ones.   For comparison, the
efficiency in finding Compton--thick AGN using other
X--ray--to--mid--infrared diagnostic ratio (e.g. L$_X$/L$_{6 micron}$) is
50\% and in an hard X--ray flux--limited survey  is about 6\%.  We have
estimated also the completeness of the method that turns out  to be of the
order of 90\%.

After having taken into account selection effects, we have estimated  the
surface density  of Compton--thick AGN down to the IRAS PSC catalogue  flux
limit (F$_{25}$= 0.5 Jy) and we have compared it with  that obtained from
an hard X--ray survey performed with SWIFT--BAT (Burlon et al.
\cite{Burlon}). We find $\rho^{CT} \sim 3 *10^{-3}$ src~deg$^{-2}$ that is
a factor 4 above the  density computed in the hard X--ray surveys. We find
that this difference is, at least in part, ascribed  to a significant
contribution ($\sim$60--90\%) of the star--forming activity to the total
25$\mu$m emission  for the sources in our sample.  By  considering only the
25$\mu$m AGN emission, we estimate a surface density  of  Compton--thick
AGN consistent  with the results found with SWIFT--BAT.

Finally, we estimate the co-moving space density of Compton--thick AGN with
L$_X$$>$10$^{43}$ erg s$^{-1}$ in a redshift range of 0.004-0.06
($\Phi$$_{C-thick}$$\sim$(3.5$^{+4.5}_{-0.5}$)$\times$10$^{-6}$ Mpc$^{-3}$).  
The prediction  for Compton--thick AGN  based on the synthesis model of the
X--ray background in Gilli et al. (2007) is consistent  with this value,
while the prediction from Treister et al. (\cite{Treistera}) is lower.

\begin{acknowledgements} 

The authors acknowledge financial support from ASI (grant n. I/088/06/0, COFIS
contract and grant n. I/009/10/0).    The authors thanks C. Vignali for
insightful suggestions and V. La Parola and G. Cusumano for useful discussion
about the BAT data. This research made use of the Simbad database and of the
NASA/IPAC Extragalactic Database (NED). We would like to thank the anonymous
referee for the useful and constructive comments which improved the quality of
the paper.

\end{acknowledgements}

\longtab{1}{
\begin{landscape}
\begin{longtable}{cccclcll}
\caption{2XMM--IRAS sample. Column 1: IRAS name; Col. 2: 2XMM name; Col. 3: offset in arcsec between the IRAS and the X-ray position;
Col. 4: NED/SIMBAD name; Col. 5: redshift (from NED/SIMBAD); Col. 6: offset in arcsec between the XMM and the optical position; Col. 7: name of the
satellite/instrument from which we have taken the X--ray data; Col. 8: X-ray classification. We
mark with ${\bf{^{**}}}$ the seven 
newly discovered Compton-thick AGN. }\\
\hline\hline
IRASname & 2XMMname & IRAS-XMM  & ID & redshift & XMM-optical & Satellite & X-ray class.\\ 
          &           & offset (") &    &          & offset (")\\ 
\hline                    
\endfirsthead
\caption{Continued.} \\
\hline\hline
IRASname & 2XMMname & IRAS-XMM  & ID & redshift & XMM-optical & Satellite & X-ray class.\\ 
      &     & offset (") &    &          & offset (")\\ 
\hline  
\endhead
\hline
\endfoot
\addtocounter{footnote}{-1}
$00085-1223$  &  J001106.5--120626 & 2.73  &  NGC 17		     &  0.0196  & 0.06   & XMM     & C-thick \\       
$00387+2513$  &  J004128.0+252958  & 11.00 &  NGC 214$^{**}$	     &  0.0151  & 0.90   & XMM+BAT & C-thick \\    
$01091-3820$  &  J011127.5--380500  & 3.00  &  NGC 424  	     &  0.0118  & 0.78   & XMM+BAT & C-thick \\       
$01306+3524$  &  J013331.1+354006  & 2.80  &  NGC 591		     &  0.0152  & 1.20   & XMM     & C-thick \\       
$01413+0205$  &  J014357.7+022059  & 10.80 &  Mrk 573		     &  0.0172  & 1.08   & XMM     & C-thick \\       
$02401-0013$  &  J024240.7--000046  & 2.63  &  NGC 1068 	     &  0.0038  & 1.80   & XMM+BAT & C-thick \\      
$03012-0117$  &  J030349.0--010613  & 2.60  &  NGC 1194 	     &  0.0136  & 0.66   & XMM+BAT & C-thick \\      
$03106-0254$  &  J031308.7--024319  & 14.30 &  2MFGC 2636$^{**}$     &  0.0272  & 0.18   & XMM     & C-thick \\       
$03222-0313$  &  iJ032448.6-030231 & 1.30  &  NGC 1320  	     &  0.0089  & 0.24   & XMM+BAT & C-thick \\       
$03317-3618$  &  J033336.3--360825  & 5.80  &  NGC 1365 	     &  0.0055  & 0.18   & XMM+BAT & C-thick \\       
$03348-3609$  &  J033646.1--355957  & 1.63  &  NGC 1386 	     &  0.0029  & 0.72   & XMM     & C-thick	  \\  
$04507+0358$  &  J045325.7+040342  & 5.30  &  CGCG 420-015$^{**}$    &  0.0294  & 1.20   & XMM+BAT+Suz. & C-thick \\
$05093-3427$  &  J051109.0--342335  & 3.80  &  ESO 362-8$^{**}$      &  0.0157  & 1.56   & XMM     & C-thick \\      
$05189-2524$  &  J052101.4--252144  & 3.89  &  IRAS 05189-2524       &  0.0426 & 0.43  & XMM+BAT & C-thin  \\   
$06097+7103$  &  J061536.2+710214  & 4.70  &  Mrk 3		     &  0.0135  & 0.60   & XMM+BAT & C-thick \\       
$06456+6054$  &  J065008.6+605044  & 2.60  &  NGC 2273  	     &  0.0061  & 0.18   & XMM     & C-thick \\      
$07379+6517$  &  iJ074241.6+651037 & 2.00  &  Mrk 78$^{**}$ 	     &  0.0372  & 0.42	 & XMM+BAT & C-thick?\footnote{This source
is classified as Compton--thick AGN by using disk--reflection models and as Compton-thin by using the toroidal model 
(in agreement also with the analysis presented in Gilli et al. 2010).}\\
$08043+3908$  &  J080741.0+390015  & 10.90 &  Mrk 622		     &  0.0232  & 0.24   & XMM     & C-thick?\footnote{This source
is classified as Compton--thin AGN by using toroidal models and as Compton-thick by using the disk--reflection  model (in agreement with the result of the
analysis presented in Guainazzi et al. 2005).}  \\
$09320+6134$  &  J093551.5+612111  & 0.90  &  UGC 05101 	     &  0.0394  & 0.66   & XMM     & C-thick \\ 
$09497-0122$  &  J095219.1--013643  & 3.70  &  Mrk 1239 	     &  0.0199  & 0.72   & XMM     & C-thin \\ 
$09585+5555$  &  J100157.8+554047  & 1.20  &  NGC 3079  	     &  0.0037  & 0.60   & XMM+BAT & C-thick \\
$11538+5524$  &  J115628.2+550732  & 6.40  &  NGC 3982  	     &  0.0037  & 2.34   & XMM     & C-thick \\	
$12540+5708$  &  J125614.2+565224  & 3.20  &  Mrk 231		     &  0.0422  & 0.72   & XMM     & C-thick \\  
$12550-2929$  &  J125744.9--294558  & 10.17 &  ESO 443-17	     &  0.0102  & 0.78   & XMM     & C-thin  \\      
$13044-2324$  &  J130705.9--234036  & 1.30  &  NGC 4968 	     &  0.0099  & 1.20   & XMM     & C-thick \\  
$13277+4727$  &  J132952.5+471144  & 15.0  &  M51		     &  0.0020  & 4.40  & XMM      & C-thick \\  
$13362+4831$  &  J133817.5+481637  & 7.14  &  Mrk 266		     &  0.0279  & 3.88  & XMM      & C-thick  \\  
$13428+5608$  &  J134442.1+555312  & 1.00  &  Mrk 273		     &  0.0378  & 0.24  & XMM+BAT  & C-thick  \\  
$13536+1836$  &  J135602.7+182218  & 1.20  &  Mrk 463		     &  0.0503  & 1.80  & XMM+BAT  & C-thin   \\     
$15084+5711$  &  J150947.0+570002  & 9.50  &  NGC 5879  	     &  0.0026  & 2.76  & XMM (Low stat.)+Chandra & C-thin     \\    
$15295+2414$  &  J153143.4+240420  & 14.00 &  3C 321		     &  0.0961  & 1.50  & XMM      & C-thick    \\  
$15327+2340$  &  J153457.3+233011  & 9.70  &  Arp220		     &  0.0181  & 2.49  & XMM      & C-thick?      \\
$15480-0344$  &  J155041.6-035318  & 8.60  &  2MASXJ15504152-0353175 &  0.0303  & 1.68  & XMM	   & C-thick  \\
$16504+0228$  &  J165258.9+022403  & 3.70  &  NGC 6240  	     &  0.0245  & 0.24  & XMM+BAT  & C-thick  \\   
$18429-6312$  &  iJ184744.1--630924 & 6.50  &  IC 4769$^{**}$	     &  0.0151  & 0.66  & XMM	   & C-thick  \\  
$19254-7245$  &  J193121.5--723920  & 4.90  &  AM 1925-724	     &  0.0617  & 2.52  & XMM	   & C-thick  \\  
$20162-5246$  &  J201958.9-523718  & 2.00  &  IC 4995		     &  0.0161  & 1.02  & XMM	   & C-thin\footnote{This source was
previously classified as Compton-thick by Guainazzi et al. 2005 and as Compton--thin by Noguchi et al. 2009.}      \\
$20305-0211$  &  J203306.1-020128  & 6.50  &  NGC 6926  	     &  0.0196  & 10.08 & XMM	   & C-thick   \\
$22045+0959$  &  J220702.0+101401  & 11.00 &  NGC 7212  	     &  0.0267  & 1.02  & XMM+BAT  & C-thick?\footnote{This source
is classified as Compton--thin AGN by using toroidal models and as Compton-thick by using the disk--reflection  model (in agreement 
with the Della Ceca et al. 2008 and references therein).}      \\
$22469-1932$  &  J224937.0-191627  & 7.00  &  MCG-03-58-007	     &  0.0315  & 1.20  & XMM+Suz. & C-thin   \\
$23024+1203$  &  J230456.6+121921  & 2.60  &  NGC 7479$^{**}$	     &  0.0079  & 0.90  & XMM+BAT  & C-thick  \\
$23156-4238$  &  J231823.5-422213  & 3.84  &  NGC 7582  	     &  0.0053  & 0.66  & XMM+BAT  & C-thick  \\
$23254+0830$  &  J232756.7+084645  & 3.50  &  NGC 7674  	     &  0.0289  & 0.72  & XMM	   & C-thick  \\
\end{longtable}
\end{landscape}
}

\end{document}